\documentclass{epl}
\usepackage{latexsym}

\title{Fractional-flux Little-Parks resistance oscillations \\ in
disordered superconducting Au$_{0.7}$In$_{0.3}$ cylinders}
\shorttitle{Fractional-flux Little-Parks resistance oscillations}
\author{Yu. Zadorozhny \and Y. Liu}
\shortauthor{Yu. Zadorozhny \etal}
\institute{Department of Physics, The Pennsylvania State University -
University Park, PA 16802, U.S.A.}

\begin{document}

\maketitle

\begin{abstract}
Resistance of disordered superconducting Au$_{0.7}$In$_{0.3}$ cylindrical
films was measured as a function of applied magnetic field.  In the
high-temperature part of the superconducting transition regime, the
resistance oscillated with a period of $h/2e$ in the unit of the enclosed
magnetic flux.  However, at lower temperatures, the resistance peaks split.
We argue that this splitting is due to the emergence of an oscillation with
a period of $h/4e$, half of the flux quantum for paired electrons.  The
possible physical origin of the $h/4e$ resistance oscillation is discussed
in the context of new minima in the free energy of a disordered
superconducting cylinder.
\end{abstract}

The essence of superconductivity lies in the long-range phase coherence,
which is manifested, in particular, in the Little-Parks (L-P) resistance
oscillation \cite{L-P}.  Strong disorder, magnetic field, or Coulomb
interaction suppresses this long-range phase coherence, leading in two
dimensions (2D) to the superconductor-to-insulator (S-I) transition
\cite{SIT}.  Several intriguing physical pictures have emerged from the
study of systems in the vicinity of the 2D S-I transition.  For example, in
disordered 2D films, Cooper pairs may form but do not condense into a
superconducting state.  These Cooper pairs then behave as (charged) point
bosons \cite{Fisher}.  Alternatively, Cooper pairs in disordered
superconducting samples can become so fragile that their non-bosonic nature
is manifested, leading, surprisingly, to negative Josephson coupling
constants \cite{S-K}.  Interestingly, a resistance oscillation with a novel
period, $h/4e$, half of the conventional flux quantum $\Phi_0=h/2e$,
emerges in these studies \cite{S-K,Glazman}.

Fractional-flux resistance oscillation (with a period of $h/2e$ rather than
$h/e$) due to coherent back scattering of unpaired electrons have been
observed in disordered {\it normal-metal} cylinders \cite{Aronov}.  Even
though the L-P resistance oscillation observed in superconducting cylinders
\cite{L-P} has the same period, $h/2e$, the flux quantum for Cooper pairs,
the physical origin of the L-P effect is very different \cite{Tinkham}.
The free energy of a superconducting cylinder is a periodic function of the
enclosed magnetic flux, resulting in an oscillation in the superconducting
critical temperature ($T_c$) and in the sample resistance.  While
superconducting rings \cite{belgium,price,reich} and cylinders
\cite{Gordon} have been studied in recent years, the effect of disorder on
the L-P resistance oscillation has never been examined systematically.  We
have carried out experimental studies of disordered superconducting Au-In
cylinders to search for oscillations with novel periods.  Here we present
experimental results obtained on Au$_{0.7}$In$_{0.3}$ cylinders for which
both $h/2e$ and $h/4e$ resistance oscillations have been observed.

To prepare a sample, a fine filament was drawn from a quartz melt and
placed across a gap in a thin glass slide.  The slide was then attached to
a rotator inside a high-vacuum evaporation system.  A cylindrical film of
Au$_{0.7}$In$_{0.3}$ was prepared by sequential deposition of 99.9999\%
pure Au and In (Au, In, Au, In, and Au in the appropriate proportion) onto
the rotating filament.  Depth profiles obtained by X-ray photoelectron
spectroscopy (XPS) and ion etching showed \cite{film} that thin alternating layers of Au and In completely inter-diffused even at room temperature, as was also found previously \cite{Simic}.
We also found that Au-In alloy adhered well to quartz without significant
oxidation, forming high quality, uniform films.  Even the thinnest
Au$_{0.7}$In$_{0.3}$ films made, a 20~nm thick cylindrical film and an 8~nm
thick planar film, exhibited low normal-state resistivity and were
superconducting.  In contrast, pure In cylinders studied in the past
\cite{MM}, as well as in our own control experiments, were not
superconducting unless a much thicker film ($\approx 300$~nm) was used.
After the deposition, the cylinder was transferred to an alumina substrate.
Several fine Au wires were attached to the cylinder using Ag epoxy, which
also anchored the cylinder to the substrate.

Electrical transport measurements were carried out in a dilution
refrigerator equipped with a superconducting magnet.  The cylinders were
manually aligned parallel to the magnetic field.  Radio-frequency filters
with insertion losses of 10~dB at 10~MHz and 50~dB at 300~MHz were used on
all electrical leads entering the sample enclosure.  The sample resistance
was measured in d.c. using a Keithley~220 current source and a Keithley~182
nanovoltmeter.  Control measurements using a manually operated battery
powered d.c. current source and also by a.c. lock-in technique yielded
identical results.  The resistance data presented here were obtained using
a bias current of 100~nA.  The reported resistance oscillations were
verified to be bias-independent throughout the Ohmic I-V regime, down to
100~pA bias current.

\begin{figure}
\onefigure[width=8.6cm]{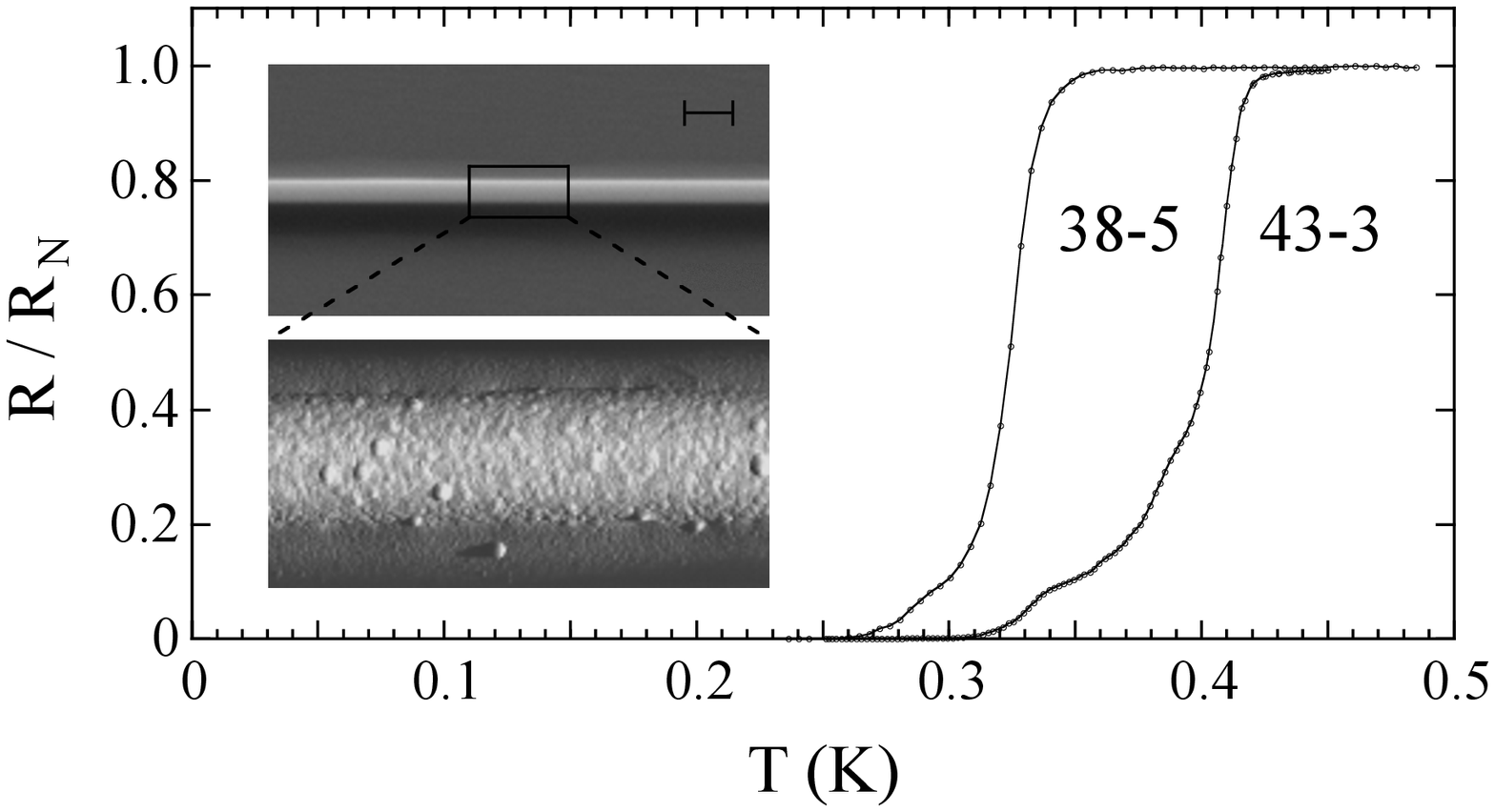}
\caption{Normalized resistances of two Au$_{0.7}$In$_{0.3}$ cylinders
plotted as a function of temperature in zero applied magnetic field.  The
insets are SEM (top) and AFM (bottom) pictures of Cylinder~43-3, capped by
5nm of Au to facilitate imaging.  The scale bar is $1~\mu$m long.  The
bottom picture is $2 \times 1~\mu$m$^2$.}
\label{1}
\end{figure}

In Fig.~\ref{1}, values of normalized resistance are plotted against
temperature for two samples, Cylinders~38-5 and 43-3.  For Cylinder~38-5,
the diameter $d=820$~nm, wall thickness $t=35$~nm, length $l=0.44$~mm, and
normal state resistance $R_N=1226~\Omega$ ($R_N^\Box=7.2~\Omega$).  For
Cylinder~43-3, $d=470$~nm, $t=30$~nm, $l=0.37$~mm, and $R_N=2248~\Omega$
($R_N^\Box=9.0~\Omega$).  The characteristic low-temperature tail in $R(T)$
was found in most of the samples and was therefore attributed to the
intrinsic properties of the material.  The onset temperature for the $h/4e$
resistance oscillation (see below) was typically above that of the tail.
We therefore conclude that the $h/4e$ oscillation was not directly related
to the presence of the tail.  We studied the film structure throughout the
entire length of several samples using a scanning electron microscope (SEM)
as well as an atomic force microscope (AFM) and confirmed that the
cylinders were granular and uniform on length scales beyond the grain size
($\approx 30$~nm).  The inset of Fig.~\ref{1} contains pictures of a
section of Cylinder~43-3 taken by SEM (large scale) and by AFM (smaller
scale).

\begin{figure}
\onefigure[width=8.6cm]{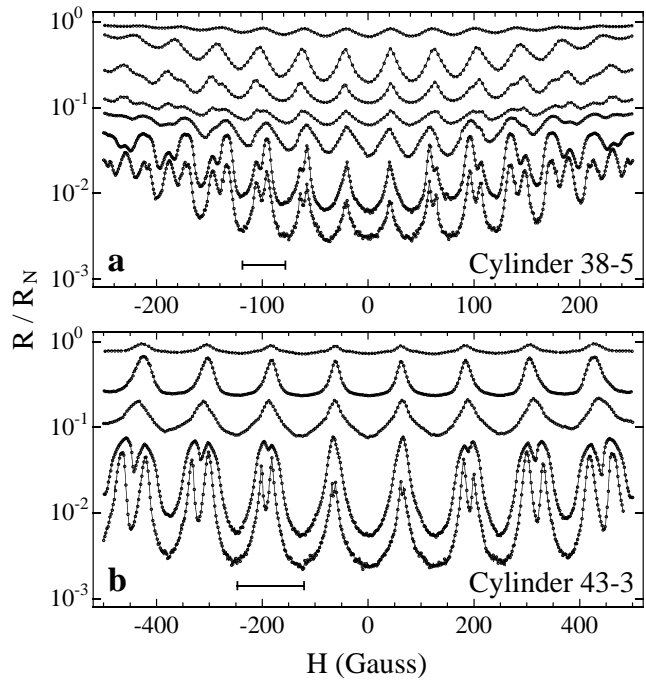}
\caption{Normalized resistance of Cylinders~38-5 ({\bf a}) and 43-3 ({\bf
b}) plotted as a function of parallel magnetic field at fixed temperatures.
>From top to bottom $T=0.33$, 0.315, 0.31, 0.297, 0.29, 0.28, and 0.27~K for
{\bf a} and 0.39, 0.36, 0.33, 0.31, and 0.30~K for {\bf b}.  The bars
indicating the oscillation periods are 40~G in {\bf a} and 125~G in {\bf
b}.}
\label{2}
\end{figure}

Fig.~\ref{2} shows normalized resistances of the two cylinders measured as
a function of magnetic field at various temperatures.  At higher
temperatures, the resistance oscillates with a single period, $\Delta H =
40$~G and 125~G for Cylinders~38-5 and 43-3 respectively, which corresponds
to an $h/2e$ oscillation.  Below certain temperature thresholds, 0.315~K
for Cylinder~38-5 and 0.320~K for Cylinder~43-3, the oscillation peaks
split as additional resistance minima develop between the minima of the
$h/2e$ oscillation.  As the temperature is lowered, the depth of these
secondary minima grows and becomes comparable to that of the primary ones.
This trend is also seen as the magnetic field is increased, with secondary
minima increasing in width and depth until the oscillation decays in higher
fields.  Similar features in resistance oscillations were found in all but
a few of more than 40 samples studied.  However, control samples of pure In
(capped with a very thin layer of Au to avoid oxidation) prepared under
similar conditions did not show any splitting of the oscillation peaks.

Can the observed splitting of the oscillation peaks be explained based on
the sample geometry?  For example, a splitting may result from coexisting
oscillations of two slightly different periods.  This could be possible if
the sample consisted of two sections of distinct diameters.  However, a
systematic
 examination of our data rules out this possibility.  If the splitting of
resistance peaks were indeed due to the presence of sections of different
diameters, we would expect to see the peaks split into three or more in
different samples or in the same sample at different temperatures.
Experimentally, the oscillation peaks always split into two.  In addition,
extensive AFM studies did not reveal any large-scale inhomogeneities or
identifiable segments of distinct diameters.

An alternative scenario is that the film might somehow consist of two
co-axial superconducting shells.  Below we show that this can also be
excluded from being the origin of the splitting.  First, as pointed out
above, sequentially evaporated Au and In completely inter-diffuse.
Therefore our samples do not consist of two coaxial cylindrical shells.
Second, even the interdiffusion were somehow incomplete, resulting in a
two-shells structure, the two shells would be coupled by proximity effect
and would behave as a single cylinder.  Our cylinders were 30nm or 35nm
thick.  The two shells would be separated by about 10nm.  However, the
zero-temperature coherence length for Au$_{0.7}$In$_{0.3}$ is 300-400nm 
(determined from the upper critical field measurements), which is much larger
than the shell separation even if they did exist.  Finally, the two-shell
scenario is inconsistent with results obtained from other samples.
Cylinders grown in a Au/In/Au sequence also showed a similar, but weaker
splitting \cite{fingerprint}.  For these samples
there is no physical basis for a two-shell scenario.  In addition, the
splitting for some cylinders (data not shown) were found to be quite large.
To account for the large splitting, the separation of the two shells would
have to exceed the film thickness, which is not possible.

The splitting of the resistance peaks can be understood if we track the
positions of the resistance minima, rather than maxima.  The primary minima
occur at magnetic field values corresponding to integer numbers of magnetic
flux quanta enclosed in the cylinder.  The positions of the secondary
minima, on the other hand, are given by half-integer numbers of flux
quanta.  The secondary minima can then be interpreted as a half-flux,
$h/4e$, resistance oscillation coexisting with the conventional $h/2e$
oscillation.  It should be emphasized that the coexistence of the $h/2e$
and $h/4e$ resistance oscillations does not necessarily imply a simple
addition of two periodic functions (see below for more discussion).

\begin{figure}
\onefigure[width=8.6cm]{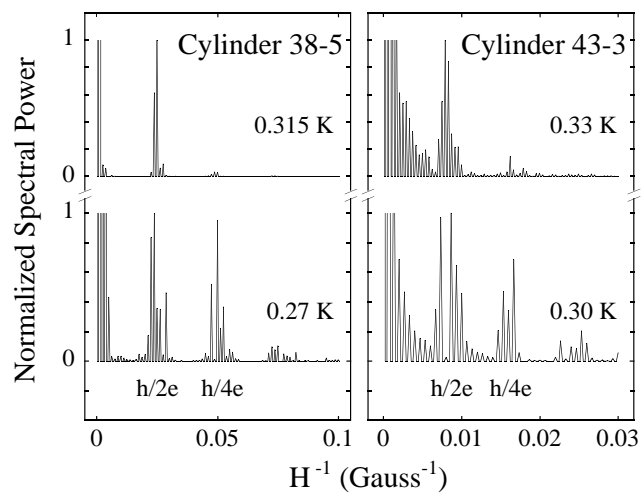}
\caption{Power spectra of selected $R(H)$ data from Figs.~\ref{2}a and
\ref{2}b.  Each spectrum is normalized to the amplitude of the $h/2e$ peak.}
\label{3}
\end{figure}

A Fourier analysis of the data was carried out.  Displayed in Fig.~\ref{3}
are Fourier power spectra of some of the $R(H)$ curves shown in
Fig.~\ref{2}.  The higher temperature spectra contain only a strong peak
around the frequency corresponding to the $h/2e$ period.  At lower
temperatures, however, a peak corresponding to an $h/4e$ oscillation
develops and eventually becomes comparable in magnitude to that of the
$h/2e$ oscillation, supporting the conclusion drawn from analyzing the
$R(H)$ data.

It has been proposed that conduction paths encircling a multi-connected
sample twice without crossing themselves would lead to an $h/4e$ resistance
oscillation \cite{solenoid}.  However, since Au$_{0.7}$In$_{0.3}$ cylinders
are structurally uniform and highly conducting even in the normal state, it
is unlikely that such non-crossing paths can exist in the sample.  An
interesting question is whether the observed $h/4e$ resistance oscillation
is indeed a consequence of the coherent back scattering of Cooper pairs as
proposed in Ref. \cite{Glazman}.  While such possibility can not be
excluded, no detailed theory of weak localization of Cooper pairs has been
worked out \cite{boson}, making it difficult to draw a conclusion on our
results in this context.

Splitting of the $h/2e$ oscillation peaks was previously observed for a
mesoscopic normal-metal (Ag) ring bordered by two superconducting (Al)
islands \cite{mirror}.  Normal electrons or holes undergo Andreev
reflection at the normal-metal\--superconductor (NS) interfaces and pick up
the phase of the superconducting order parameter.  The splitting was
believed to result from the ring's conductance being a periodic function of
the phase difference between the two superconductors with a period of
$\pi$.  Subsequently, better controlled experiments in similar geometry
demonstrated a $2\pi$ periodicity \cite{Lambert}, as expected \cite{ZS}.
An explanation for the result of \cite{mirror} is in fact still lacking.
As we argue below, the physics involved in our samples is different from
that in NS heterostructures.

Splitting was also found in the first oscillation peak for mesoscopic Al
rings \cite{belgium}.  This splitting, however, was associated with the
presence of the so-called resistance anomaly, a resistance peak in $R(T)$
just below the $T_c$.  The anomaly results from charge imbalance near an NS
interface \cite{belgium}.  No such resistance anomaly was observed in our
Au$_{0.7}$In$_{0.3}$ cylinders.  Therefore we believe that the splitting in
Al loops is unrelated to our observations.

Physically, it might be reasonable to expect that the fractional-flux
resistance oscillation can be understood on the same footing as the
conventional L-P effect.  Within the Ginzburg-Landau theory, the phase
boundary between the superconducting and the normal phases, $T_c(H)$,
obtained by tracking a fixed resistance value on the $R(H,T)$ surface,
reflects the landscape of the free energy in the $H$-$T$ space
\cite{Tinkham}.  For pure samples, for which $R(T)$ shifts rigidly as the
magnetic field is swept \cite{MM}, any resistance value can be chosen for
this purpose, as long as it is reasonably below $R_N$.

\begin{figure}
\onefigure[width=13cm]{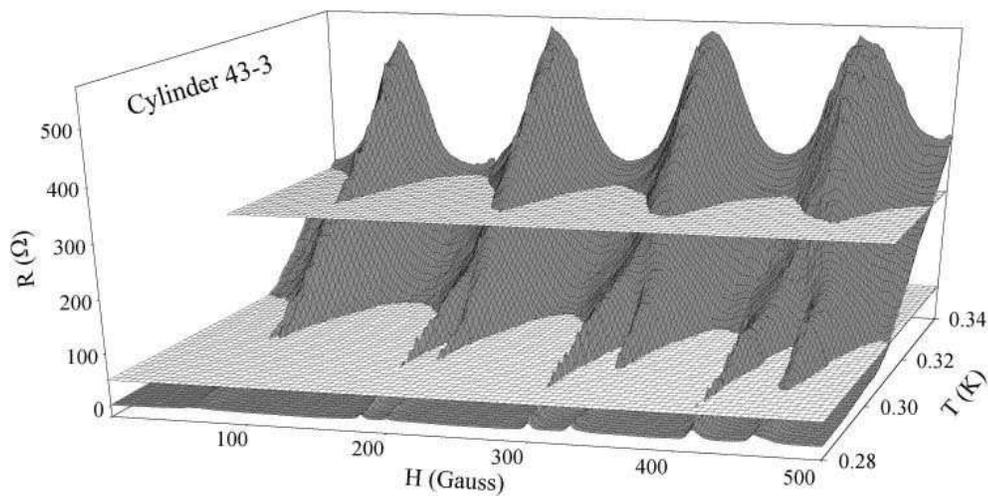}
\caption{Resistance of Cylinder~43-3 plotted as a function of temperature
and magnetic field.  "Phase boundaries" sliced out by horizontal planes at
$50~\Omega$ and $250~\Omega$ are shown (see text).}
\label{4}
\end{figure}

We have studied in detail the evolution of the cylinder resistance as a
function of $H$ and $T$, as shown in Fig.~\ref{4} for Cylinder~43-3.  It is
evident from Fig.~\ref{4} that the fractional-flux resistance oscillation
persists to the lowest temperatures at which a non-zero resistance is still
observable.  In addition, as has been mentioned, the $h/4e$ oscillation is
bias-independent for at least three decades of bias current.  These
observations suggest that the fractional periodicity is an equilibrium
property of the global superconducting state.

By analogy with the standard approach, we have obtained two "phase
boundaries" by tracking fixed resistance values of $R = 250~\Omega$ and
$50~\Omega$.  While the former curve resembles the phase boundary of a pure
superconducting cylinder, the one at 50~$\Omega$ shows splitting of the
oscillation peaks (Fig.~\ref{4}).  The $T_c(H)$ curve corresponding to $R =
50$~$\Omega$, being lower in temperature, should be closer to the landscape
of the free energy of the superconducting state of the sample.  Therefore,
the presence of the secondary resistance minima is an indication that the
free energy of Au$_{0.7}$In$_{0.3}$ cylinders develops additional minima at
half-integer numbers of flux quanta.

In the free energy consideration, it is natural to analyze the oscillation
based on the positions of the energy minima rather than maxima.
Furthermore, the free energy oscillations of $h/2e$ and $h/4e$ periods are
not simply added together, but rather the lowest energy state is chosen at
each value of the magnetic field.  This explains why no new features
appeared at the primary resistance minima as the secondary minima
developed.  It has been shown, based on microscopic consideration, that the
free energy of a non-disordered superconducting cylinder is minimized for
integer numbers of flux quanta \cite{Byers-Yang}.  The same is true for
a ring or cylinder with (conventional) Josephson junctions embedded along
the circumference.

However, the situation is different for samples containing junctions with
negative instead of positive Josephson coupling, known as $\pi$-junctions
\cite{russians}.  The free energy of a superconducting ring interrupted by
such a junction has a minimum at $\Phi_0/2=h/4e$ \cite{russians}.  Since
the free energy has to be a periodic function of the enclosed flux
\cite{Byers-Yang}, this means that free energy minima occur at all
half-integer numbers of flux quanta.  This result can be generalized to a
ring containing an odd number of $\pi$-junctions \cite{Zhou}.  (The free
energy of a ring with an even number of $\pi$-junctions has minima at
integer numbers of flux quanta, as usual.)  The free energy of a
superconducting cylinder, which can be considered as a collection of rings,
should have minima at both integer- and half-integer flux quanta if there
are odd numbers of $\pi$-junctions in some of the rings.

The above picture relies on the presence of Josephson junctions in our
cylinders.  For Au-In alloy, the maximum solid solubility of In in Au is
about 10\% \cite{alloy}.  In a Au$_{0.7}$In$_{0.3}$ film, In-rich regions
of varying size are expected to form in a uniform Au$_{0.9}$In$_{0.1}$
matrix as the excess In precipitates.  Such phase separation has been
confirmed by XPS studies of planar Au$_{0.7}$In$_{0.3}$ films \cite{film}
and appears to occur in cylindrical films as well.  The characteristic
length scale on which the In concentration varies is given by the average
grain size.  At low temperatures, superconducting In-rich regions are
linked by Josephson coupling, forming a random superconductor-normal
metal-superconductor (SNS) junction array.

Negative Josephson coupling in these SNS Josephson junctions may emerge
from mesoscopic fluctuations in disordered superconductors, as proposed
originally by Spivak and Kivelson \cite{S-K} for films close to the S-I
transition in zero magnetic field.  Zhou and Spivak \cite{Zhou1}
have analyzed disordered superconducting films in a parallel field, a
case applicable to our system.  In particular, the Josephson coupling
between two superconducting droplets embedded in a normal metal in parallel
magnetic field, and its fluctuation, have been calculated explicitly
\cite{Zhou2}.  As the magnetic field increases, the average coupling
vanishes while the fluctuation in the coupling remains essentially
unchanged.  This is only possible if both positive and negative couplings
are allowed.  Otherwise, the fluctuation would have to go to zero along
with the average.  Therefore, the pi-junction becomes more favored as the
field increases, leading to an increase in the part of free energy that
possesses minima at half-integer flux quanta.  This will in turn enhance
the h/4e resistance oscillation, which may account for the observed
increase in the $h/4e$ oscillation component in higher fields.

In conclusion, we have observed a splitting of peaks in the resistance
oscillation in Au$_{0.7}$In$_{0.3}$ cylinders close to the bottom of the
superconducting transition.  We have argued that this is due to the
emergence of a resistance oscillation of period $h/4e$.  We have examined
this oscillation in the context of the presence of new minima in the free
energy of the cylinders at half-integer numbers of flux quanta.  More
theoretical work is needed to clarify the physical origin of the observed
$h/4e$ resistance oscillation.

\acknowledgments
The authors would like to acknowledge useful discussion with A.~M.~Goldman,
J.~K.~Jain, S.~Kivelson, M.~Ma, and F.~Zhou.  They would also like to
acknowledge valuable contributions of David Herman and Karl Nelson to this
project.  This work is supported by the NSF under grant DMR-9702661.

\break

\end{document}